\documentclass[superscriptaddress,twocolumn,prb,showpacs,preprintnumbers,amsmath,amssymb]{revtex4}
\usepackage{graphicx}
\usepackage{color}
\usepackage{ulem}
\begin{document}

\title{Inhomogeneous magnetism in the doped kagome lattice of LaCuO$_{2.66}$}

\author{M.-H. Julien}
\email{marc-henri.julien@lncmi.cnrs.fr}
\affiliation{Laboratoire National des Champs Magn\'etiques Intenses, CNRS-UJF-UPS-INSA, 38042 Grenoble Cedex 9, France}

\author{V. Simonet}
\email{virginie.simonet@grenoble.cnrs.fr}
\affiliation{Institut N\'eel, CNRS-UJF, 25 avenue des Martyrs,  38042 Grenoble Cedex 9, France}

\author{B. Canals}
\affiliation{Institut N\'eel, CNRS-UJF, 25 avenue des Martyrs, 38042 Grenoble Cedex 9, France}

\author{R.~Ballou}
\affiliation{Institut N\'eel, CNRS-UJF, 25 avenue des Martyrs, 38042 Grenoble Cedex 9, France}

\author{A.K.~Hassan}
\affiliation{Laboratoire National des Champs Magn\'etiques Intenses, CNRS-UJF-UPS-INSA, 38042 Grenoble Cedex 9, France}

\author{M.~Affronte}
\affiliation{CNR NANO S3 and Dipartimento di Fisica, Universit‡ di Modena e Reggio Emilia, via G. Campi 213A 41125, Modena, Italy}

\author{V.O.~Garlea}
\affiliation{Institut N\'eel, CNRS-UJF, 25 avenue des Martyrs, 38042 Grenoble Cedex 9, France}
\affiliation{Oak Ridge National Laboratory, 1 Bethel Valley Road, PO Box 2008, Oak Ridge, Tennessee 37831}

\author{C.~Darie}
\affiliation{Institut N\'eel, CNRS-UJF, 25 avenue des Martyrs, 38042 Grenoble Cedex 9, France}

\author{P.~Bordet}
\affiliation{Institut N\'eel, CNRS-UJF, 25 avenue des Martyrs, 38042 Grenoble Cedex 9, France}

\date{\today}

\begin{abstract}

The hole-doped kagome lattice of Cu$^{2+}$ ions in LaCuO$_{2.66}$ was investigated by nuclear quadrupole resonance (NQR), electron spin resonance (ESR), electrical resistivity, bulk magnetization and specific heat measurements. For temperatures above $\sim180$~K, the spin and charge properties show an activated behavior suggestive of a narrow-gap semiconductor. At lower temperatures, the results indicate an insulating ground state which may or may not be charge ordered. While the frustrated spins in remaining patches of the original kagome lattice might not be directly detected here, the observation of coexisting non-magnetic sites, free spins and frozen moments reveals an intrinsically inhomogeneous magnetism. Numerical simulations of a 1/3-diluted kagome lattice rationalize this magnetic state in terms of a heterogeneous distribution of cluster sizes and morphologies near the site-percolation threshold.

\end{abstract}
\maketitle

\section{Introduction}

\begin{figure}
\centerline{\includegraphics[width=3.2in]{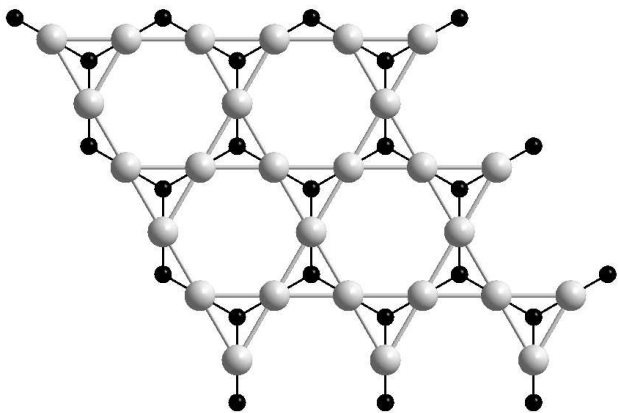}}
\caption{Copper (large light spheres) and oxygen (small dark spheres) sites in the CuO$_{0.66}$ kagome planes of LaCuO$_{2.66}$.\cite{Garlea03} These planes alternate with layers of LaO$_6$ octahedra.} 
\label{structure}
\end{figure}

Whether the physics of quantum spin liquids is relevant to the occurrence of high temperature superconductivity in copper oxides has been one of the most debated issues in condensed matter physics over the last two decades. This question has triggered important research efforts in two directions: i) the physical realizations of quantum spin liquids, and ii) the doping of such systems with charge carriers. Natural candidates for realizing superconductivity from a spin liquid are compounds such as transition-metal oxides where unpaired electrons reside on dimer, ladder or triangle-based networks.\cite{Reviews} In general, two-dimensional lattices are more favorable than 1D systems for which localization is expected at low temperature ($T$).

Copper oxides RCuO$_{2+\delta}$ (R = Y, La, In, Sc, Nd)~\cite{Cava93,Isawa97,Garlea03,delafossitenew} with a delafossite structure~\cite{delafossite} provide an interesting playground for studying the effect of doping in a 2D lattice where Cu ions form triangles. Depending on the oxygen content $\delta$, a variety of triangle-based magnetic networks can be obtained as the additional oxygen O$_\delta$ occupies different positions in the Cu planes ({\it i.e.} the geometry of the magnetic network and the formal Cu valence are necessarily linked together). Substantial progress in the synthesis of these materials has recently led to the identification of well-defined structures.\cite{delafossitenew}

Among copper delafossites, LaCuO$_{2.66}$ appears to be of particular interest. It consists of layers of edge-sharing LaO$_6$ octahedra alternating with triangular planes of Cu$^{2+}$ ions. With respect to LaCuO$_2$, the extra oxygen ions are located in the Cu planes. For the concentration $\delta=2/3$, the oxygen ordering is such that the magnetic network connecting the Cu ions has a perfect kagome geometry, made of corner sharing equilateral triangles with an oxygen at their center (Fig.~\ref{structure}).\cite{Garlea03} This ideal situation should be contrasted with that in the almost isostructural compound YCuO$_{2.66}$ which will be mentioned several times in this paper. The CuO$_{0.66}$ planes of this yttrium counterpart also form a kagome lattice but the distorted structure probably leads to non-negligible next-nearest neighbor ({\it i.e.} inter-triangle) exchange interactions.\cite{LeBacq05} The highly frustrated kagome lattice, of which there exists a number of good, but never perfect, experimental realizations (see refs.~\cite{SCGO,volborthite,herbertsmithite} for just a few examples) is one of the most popular candidate for a quantum spin liquid in two dimensions.

With a formal valence of Cu ions of +2.33, LaCuO$_{2.66}$ (as well as YCuO$_{2.66}$) should be considered either as a kagome lattice doped with 33~\% of itinerant holes or as a diluted kagome lattice with 33~\% of non-magnetic (Cu$^{3+}$) sites (Note that band structure calculations indicate that the states at the Fermi level have predominantly Cu $d_{z^2-r^2}$ character~\cite{Mattheiss93}). Both possibilities are of considerable theoretical interest.\cite{hubbard,impurities} Although LaCuO$_{2.66}$ has received early theoretical attention,\cite{Mattheiss93,Dolores96} little is known concerning its electronic properties. For $\delta=0.64$, which can be viewed as a disordered version ({\it i.e.} with 0.02 oxygen vacancies) of the stoichiometric $\delta= 0.66$, early studies found a metallic ground state with strong correlations and strong disorder effects close to the metal-insulator transition.\cite{Cava93,Ramirez94,Walstedt94} For the more recently synthesized $\delta=0.66$ concentration, which presents improved structural homogeneity, a muon spin rotation ($\mu$SR) study found antiferromagnetic (AFM) order in $\sim$20\% of the sample volume below 40-50~K.\cite{Mendels04}

The paper is organized as follows: Sections II, III and IV are devoted to experimental results in LaCuO$_{2.66}$. Section V presents numerical simulations of the kagome lattice at a 1/3 dilution level, a description justified by the experimental evidence that the doped holes are strongly localized at low $T$. Section VI summarizes the results. Key issues and open questions are discussed in section VII.

\section{Experimental details}

Polycrystalline samples were used for all measurements. They were prepared by solid state reaction followed by an oxidization procedure with long annealing times at the lowest possible temperatures in order to get good quality samples.\cite{Garlea03} The results from magnetization and specific heat measurements showed minor sample dependence at the quantitative level, presumably due to the different amount of impurities, but not at the qualitative level. The magnetization data shown in this article were obtained on the sample used for NQR measurements and for a $\mu$SR study.\cite{Mendels04}

The specific heat was measured by the pulsed relaxation method and the resistivity was measured by the standard four-contact method, both in a commercial setup (Quantum Design PPMS). The magnetization was measured by the axial extraction method in a home-built magnetometer in the temperature range 2-800 K and in magnetic fields up to 10 T as well as on a Quantum Design SQUID magnetometer (MPMS) in the range 2-300~K and in fields up to 5~T.

The ESR measurements were performed at different frequencies and temperatures using a home built multifrequency high-field spectrometer. Different frequencies were investigated (54$-$115~GHz) using back-wave oscillators and Gunn diode sources, guided to the sample with oversized waveguides in a transmission probe. The spectra were recorded as a function
of magnetic field at a fixed frequency.

The $^{139}$La and $^{63,65}$Cu NQR measurements were performed on a home-built pulsed spectrometer. Let us recall that NQR measurements, unlike nuclear magnetic resonance (NMR), are performed in zero external magnetic field. The signal intensity was determined from the integral over time of the spin-echo signal, and divided by $f^2$ as a function of NQR frequency $f$. The spin-lattice relaxation time ($T_1$) was estimated from a single-exponential fit of the time dependence of the nuclear magnetization $M_z$ after a single saturation pulse: $M_z(t)=M(t=\infty)(1-\exp(-(3t/T_1)^\alpha))$.~\cite{Walstedt} The stretching exponent $\alpha$ was found to vary smoothly from 0.9 at 300~K to 0.7 at 4.2~K, indicating a moderate distribution of $T_1$ values.~\cite{Johnston06,Mitrovic08}

\section{Macroscopic measurements}

\begin{figure}[t!]
\centerline{\includegraphics[width=3.5in]{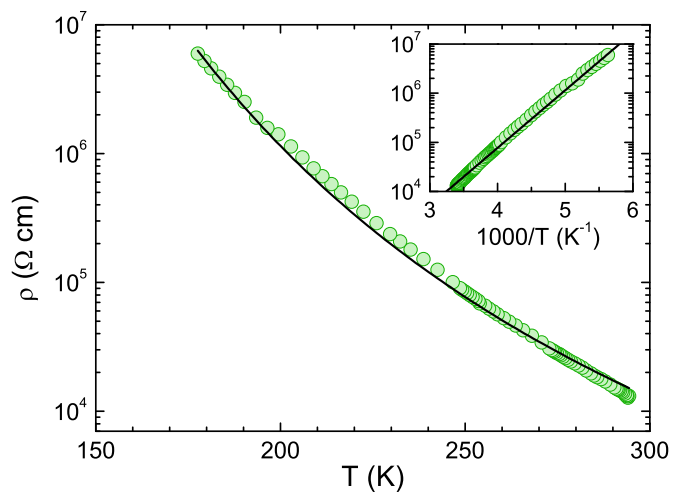}}
\caption{(Color online) Main panel: electrical resistivity {\it vs.} temperature. The line represents a thermally activated behavior (see text). (Inset) Same data {\it vs.} inverse temperature. }
\label{resistivity}
\end{figure}

\subsection{Resistivity}

The resistance rapidly increases on cooling and becomes impossible to measure reliably below $\sim$150 K, most probably because of a large contact resistance. The intrinsic transport properties of this oxide are thus difficult to determine from bulk measurements on a pressed pellet. As previously reported in LaCuO$_{2.64}$ \cite{Cava93}, the resistance of LaCuO$_{2.66}$ does not follow a simple activated behavior over the whole temperature range (Fig.~\ref{resistivity}). A fit to an activated law $\rho=\rho_0\exp(\Delta_R/k_BT)$ can nonetheless be performed over limited temperature intervals between 300 and 180 K. This yields an activation energy $\Delta_\rho=2700\pm300$~K. Although the $T$ dependence may be affected or even dominated by the grain boundaries, an activated law is also found in other measurements (see below) and is thus consistent with a semi-conducting behavior.

\subsection{Bulk magnetization}

\begin{figure}[t!]
\centerline{\includegraphics[width=3.1in]{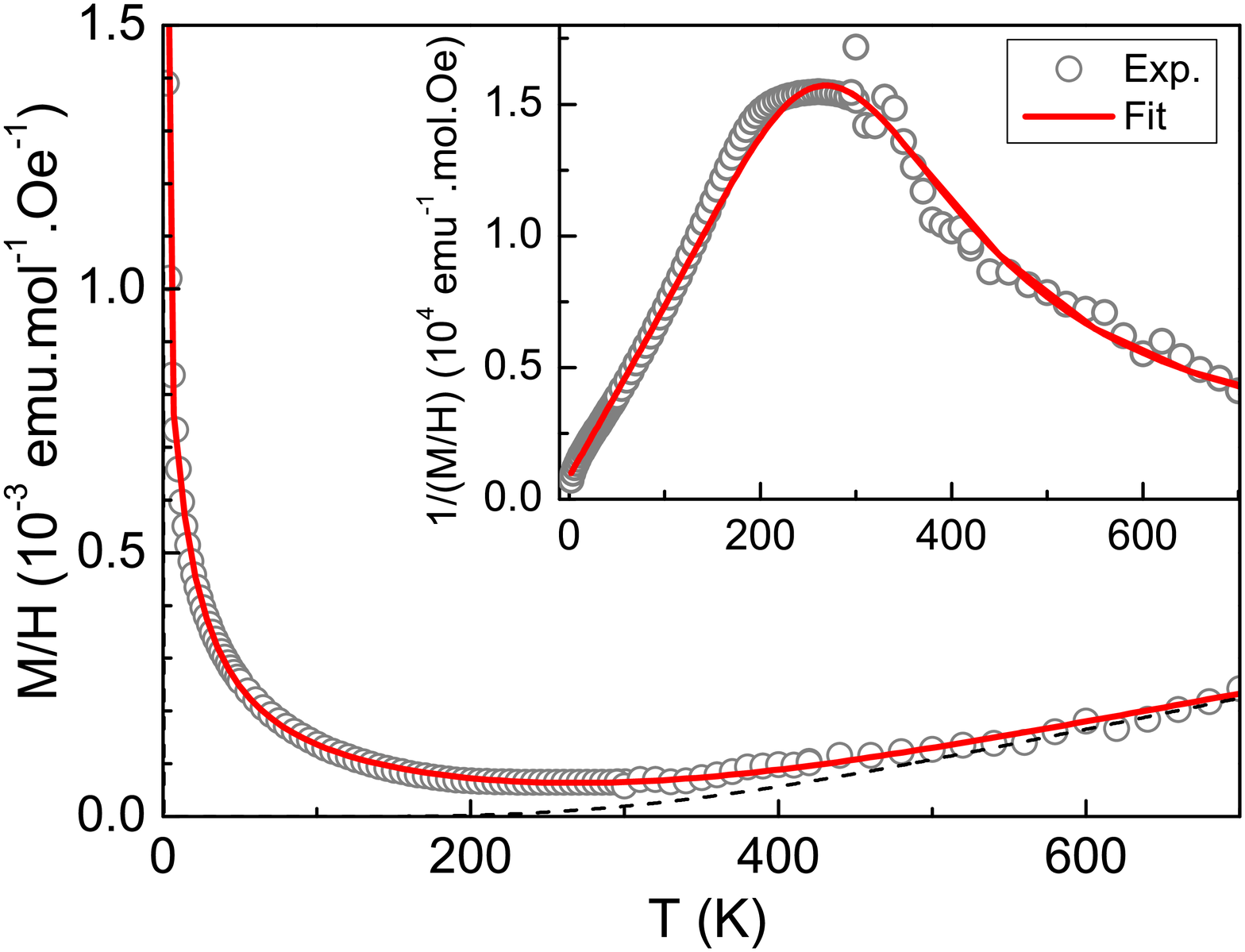}}\caption{(Color online) $M/H$ {\it vs.} temperature $T$. Data above and below 300~K results from two separate measurements: in the temperature range [2-300 K], $M$ was measured in a field $H$=10 kOe while in the range [300-700 K], $M/H$ was extracted from linear fits of $M$ versus $H$ up to 60 kOe. The two methods give the same value of $M/H$ at 300~K. The line is a fit following Eqn.~(1). Dashes represent the function $a\times\exp(-\Delta_{\chi}/k_BT)$ with $\Delta$=1281~K. The inverse curves are displayed in the inset.}
\label{suscep}
\vspace{5mm}
\centerline{\includegraphics[width=3.1in]{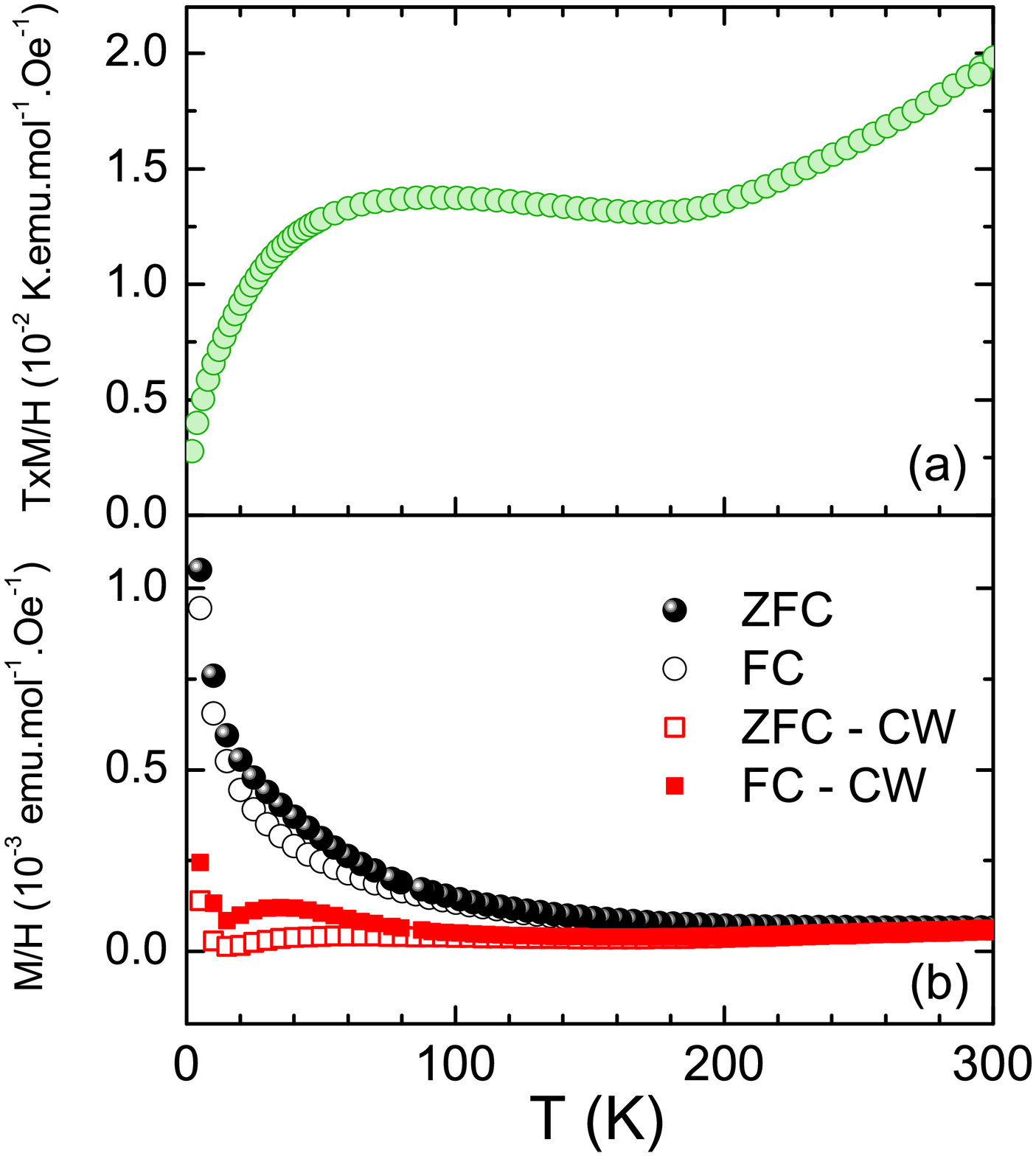}}\caption{(Color online) (a): $M/H$ times $T$ versus $T$, in $H$=10~kOe. (b) $M/H$ {\it vs.} $T$ measured in 100~Oe after zero-field cooling (open circles) and after cooling in a field of 100 Oe (filled circles). Open and filled squares are the same data from which the Curie-Weiss contribution $\chi_0+C/(T+\theta)$ has been subtracted (see text). The residual upturn below $\approx$~13~K is possibly due to paramagnetic impurities.}
\label{ZFC}
\end{figure}

The bulk magnetization $M$ was found temperature reversible up to 750~K, above which the sample is subject to chemical transformation. $M/H$ is equal to the linear susceptibility $\chi$ down to 5 K, below which the isothermal magnetization versus field starts to deviate from linearity. As can be seen in Fig.~\ref{suscep}, $M/H$ displays a non monotonic temperature dependence. The $T\times M/H$ versus $T$ representation of Fig.~\ref{ZFC} gives another view on the different regimes. The data suggest the presence of two contributions with opposite $T$ evolutions. Starting from high temperatures, the decrease of $M/H$ on cooling is tentatively modeled by an activated behavior. Then, the sharp rise of $M/H$ below $\sim$200~K can be accounted for by a Curie-Weiss law. The whole susceptibility then writes:

\begin{equation}
\chi=\chi_0+C/(T+\theta)+a\times\exp(-\Delta_{\chi}/k_BT) ,
\end{equation}
and the following values are obtained from fitting of the data to this expression: a constant term due to sample environment and diamagnetic contribution $\chi_0=-4.5\times 10^{-6}$ emu/(mol.Oe), the Curie constant $C=0.016 \pm 0.002$ emu.K/(mol.Oe), the Curie-Weiss temperature $\theta=13.8\pm 5$ K, $a=1.34\times 10^{-3}$ emu/(mol.Oe), and the susceptibility gap $\Delta_{\chi}/k_B=1281\pm 200$~K.

If a gap were present in the spin excitations, its maximum possible magnitude would be set by the main magnetic exchange coupling $J$. The intra-triangle $J$ in LaCuO$_{2.66}$ is not known but it is expected to be of the order of a few hundred Kelvin, since $J$ values range from 170 to 660~K in YCuO$_{2.5}$, a complex two dimensional array of similar Cu$^{2+}$ triangles with an oxygen at their center.\cite{LeBacq05,Capponi} Inter-plane magnetic couplings, on the other hand, are considered to be much weaker than in-plane couplings, despite the d$_{z^2-r^2}$ orbital occupation.\cite{LeBacq05} It is thus unlikely that the large gap magnitude $\Delta_{\chi}\sim 10^3$~K can be explained by a spin gap. Instead, the similarity in the orders of magnitude of the gap values suggests that the activated behaviors in the resistivity and susceptibility have a common origin. The simplest scenario we can think of is a thermal activation of states in the valence band of a small-gap semiconductor (see for instance ref.~\cite{semiconductors}). The rise of $\chi$ at high temperatures then results from the increasing contribution of the Pauli susceptibility of the conduction electrons.

The magnitude of the Curie constant $C$ indicates that the Curie-Weiss behavior of the susceptibility arises from 5\% of spins 1/2 (relative to the total number of Cu ions). This term will be shown below to be mostly of intrinsic origin. The fitting, however, does not capture fine details of the shape of $M/H$, especially the non linearity of the inverse of $M/H$ at low $T$ (inset of Fig.~\ref{suscep}). This can be due to the contribution from collective spins 1/2 from small clusters with a complex geometry~\cite{Robert,Azuma} and/or to the onset of partial magnetic ordering whose existence is supported by the slight difference between field-cooling and zero-field-cooling below $\approx$80 K. This, together with a progressive downturn of $T\times M/H$ on cooling (Fig.~\ref{ZFC}) indicates that the viscosity of the spin correlations progressively increases due to some sort of freezing in the spin system. However, the relatively modest amplitude of the effects observed in the susceptibility (in particular as compared to YCuO$_{2.66}$~\cite{Simonet}) is inconsistent with a conventional transition towards bulk long range order. This observation agrees with a spin freezing in only $\sim$20\% of the sample volume as inferred from $\mu$SR.\cite{Mendels04}

\begin{figure}
\centerline{\includegraphics[width=3.4in]{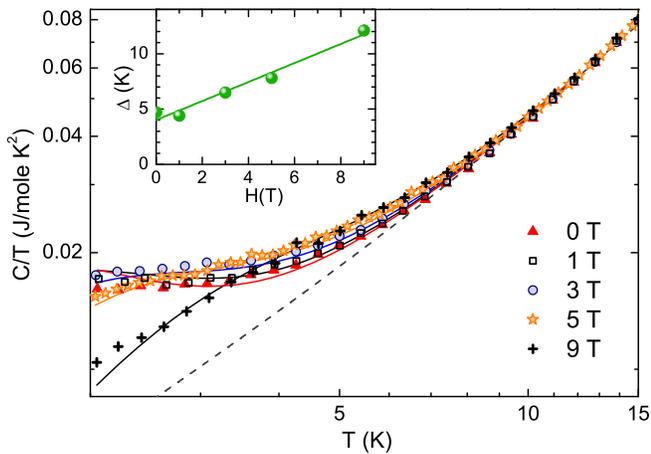}}
\caption{(Color online) specific heat $C_p$ on a log-log scale at different magnetic field values. Solid lines are fits of the data (symbols) with $C_p=\beta T^3+\beta_2 T^2+\alpha R({\Delta\over k_bT})^2\exp{-({\Delta\over k_bT})}/(1+e^{-{\Delta\over k_bT}})^2$ with $\beta$=0.00015~J.K.mol$^{-1}$, $\beta_2$=0.00294~J.mol$^{-1}$, $\alpha$=0.0064 J/mole K$^2$ and the values of $\Delta$ are shown in the inset as a function of magnetic field. The dashed line shows the $T^3 +T^2$ contribution.}
\label{specific}
\end{figure}

\subsection{Specific heat}
\begin{figure*}[t!]
\centerline{\includegraphics[width=7in]{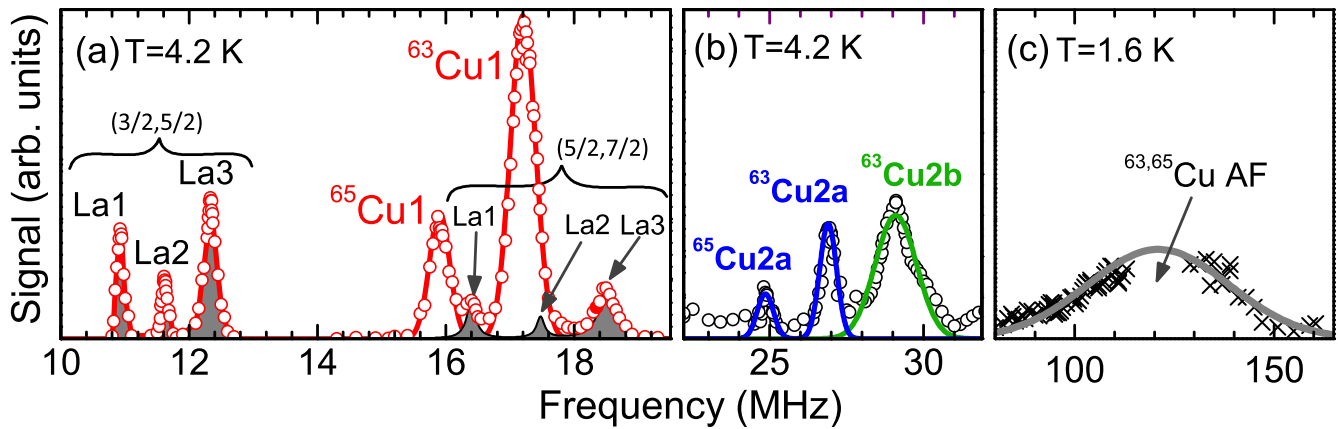}}
\caption{(Color online) NQR spectrum of LaCuO$_{2.66}$ at $T=4.2$~K. (a) Three different $^{139}$La sites (grey peaks) and a main $^{63,65}$Cu1 site are resolved. The three La sites are observed in two different sets of three lines, that correspond to the $\pm\frac{3}{2}\leftrightarrow\pm\frac{5}{2}$ and $\pm\frac{5}{2}\leftrightarrow\pm\frac{7}{2}$ transitions within the energy levels of the nuclear spin $^{139}I=\frac{7}{2}$. The $\pm\frac{1}{2}\leftrightarrow\pm\frac{3}{2}$ transitions occur at too low frequency to be observed with our setup. Both $^{63}$Cu and $^{65}$Cu have a spin $I=3/2$, thus only one $\pm\frac{1}{2}\leftrightarrow\frac{3}{2}$ NQR transition for each isotope/site. (b) Cu2a and Cu2b sites corresponding to minority paramagnetic sites (see text). The total intensity of these sites represents a few \% of the Cu1 intensity. (c) At $T=1.6$~K, the broad $^{63,65}$Cu resonance around 121 MHz is attributed to sites in presence of a 10.2~T internal field, {\it i.e.} magnetic order. The absence of data points around the peak maximum is due to the presence of a spurious signal from the NQR probe.}
\label{spectra}
\end{figure*}
At high temperature, the specific heat $C_p$ measured in zero magnetic field approaches the expected Dulong and Petit limit. No sharp transition is detected in the investigated temperature range, thereby ruling out the presence of a conventional, homogeneous, long-range magnetic order (see Fig.~\ref{specific}). A field dependence is observed below 10~K. No evidence for an activated behavior, that could reveal a large spin gap, was found in these data. The data below 11~K were fitted by the sum of several contributions (see Fig.~\ref{specific}):

- a $T^3$ term accounting for the Debye lattice, which cannot be evaluated independently here because there is no non-magnetic isostructural reference compound.

- a $T^2$ term, as previously reported for LaCuO$_{2.64}$.\cite{Ramirez94} The $T^2$ term in LaCuO$_{2.66}$ is most probably of magnetic origin since we found it to be absent in the specific heat of LaCuO$_2$ which, although not isostructural with LaCuO$_{2.66}$, presents a similar lamellar structure. Such a $T^2$ law has been observed for instance in the kagome bilayer of SCGO even at large dilution levels~\cite{Ramirez92} and has received theoretical interpretations in terms of excitations of various spin liquid phases of the $S=1/2$ Heisenberg model on the kagome lattice.\cite{Ryu,Ran} It could also arise from excitations of the 2D-AFM order.

- a Schottky anomaly, likely due to the presence of discrete low energy levels, accounts for the field dependence of $C_p$. The deduced energy barrier varies linearly with the field (inset to Fig.~\ref{specific}). Assuming that this terms arises only from two-level paramagnetic impurities, it would account for 0.6 \% of the magnetic Cu. However, the observation of a residual contribution in zero field reveals that the Schottky term does not solely originate from impurities but also from the distribution of discrete energy levels characterizing the small clusters.

- a temperature-linear term was tentatively introduced. However, we could clearly better fit our data without this last term. Since this term is attributed to conduction electrons in LaCuO$_{2.64}$~\cite{Ramirez94}, its absence here is another indication that LaCuO$_{2.66}$ has an insulating ground-state.

\section{Local probe measurements}

\subsection{$^{139}$La NQR}

Starting from its low frequency side, the NQR spectrum of LaCuO$_{2.66}$ at $T=4.2$~K (Fig.~\ref{spectra}) first shows that there are three different $^{139}$La sites from the electric field gradient point of view. The relative intensities of these La1, La2 and La3 sites represent approximately 27, 13 and
60~\% of the total La intensity. In principle, there are six different crystallographic sites for La$^{3+}$ ions in this material.\cite{Garlea03} Nevertheless, these six sites can be grouped into three categories, according to the number (six, seven or eight) of oxygen anions to which each La$^{3+}$ is coordinated.\cite{Garlea03} The relative population of these three types of coordination cannot be accurately determined because of uncertainties regarding the complex stacking sequence of copper-oxygen planes and the occurrence of numerous stacking faults.\cite{Garlea03} Still, assuming an equal probability for the three possible stacking sequences (named A, B and C in Ref.~\cite{Garlea03}), the sixfold, sevenfold and eightfold coordinated La would represent 12, 44 and 44~\% of the sites, respectively. Given the large uncertainties, these numbers appear compatible with the relative NQR intensities so that the three different $^{139}$La NQR sites could indeed correspond to the three different coordination numbers of La$^{3+}$ ions. It cannot be excluded, however, that part of the differentiation of electric field gradients at the La site is related to the distribution of charges in CuO$_\delta$ planes.

\subsection{$^{63,65}$Cu1 NQR}

\begin{figure}
\centerline{\includegraphics[width=3.5in]{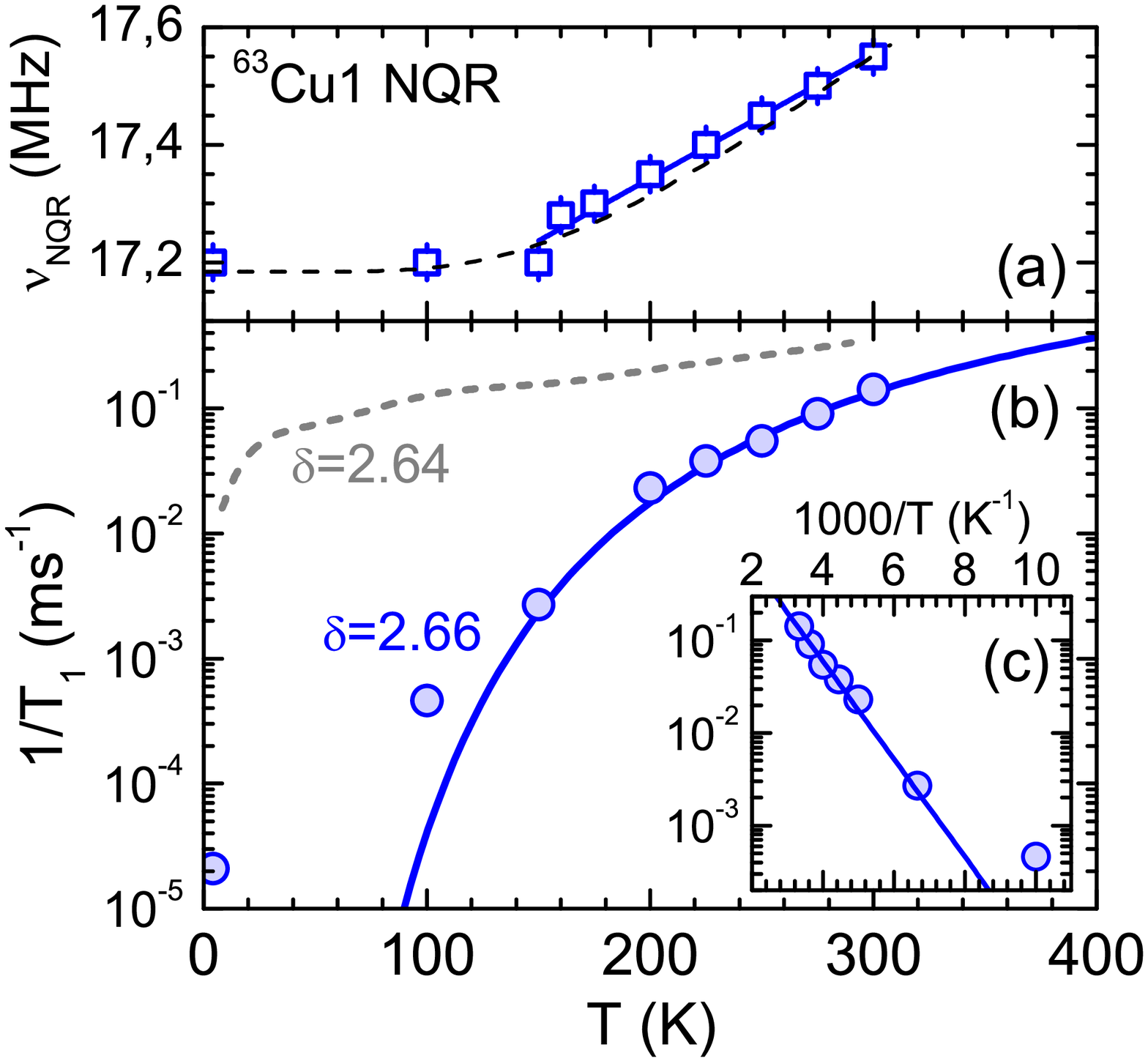}} \caption{(Color online) (a) $T$ dependence of the NQR frequency of the Cu1 site, with a clear saturation below 150 K. The continuous (blue) line is a linear fit to the data from 150~K to 300~K. The dashed (black) line is an exponential fit $a+b\exp(-E/k_BT)$ with $E=622$~K. (b) $T$ dependence of $1/T_1$ for the Cu1 site, compared to data in LaCuO$_{2.64}$ from ref.~\cite{Walstedt94}. The solid line is a fit to an activated behavior $\exp(-\Delta_{T_1}/k_BT)$ with $\Delta_{T_1}=1216$~K, which holds from 300~K down to approximately 150~K. (c) Same $1/T_1$ data and fit in log scale {\it vs.} $T^{-1}$.}
\label{cu1nqr}
\end{figure}

Around 16-17 MHz, a single, intense, NQR line is observed for each of the two isotopes $^{63}$Cu and $^{65}$Cu. This NQR site is named Cu1 in Fig.~\ref{spectra}. From their $^{63}$Cu NMR line in LaCuO$_{2.64}$, Walstedt {\it et al.} deduced a quadrupole coupling constant $\nu_Q=$~15.5~MHz and an asymmetry parameter of the electric field gradient tensor $\eta=0.65$ at $T=293$~K.\cite{Walstedt94} These parameters imply a NQR frequency $\nu_{\rm NQR}=\nu_Q \sqrt{1+\eta^2/3}=17.7$~MHz quite close to our measured value of 17.55~MHz at 300~K (Fig.~\ref{cu1nqr}a). This suggests that the Cu1 site in LaCuO$_{2.66}$ could have a similar electronic environment {\it at high temperature} as the site observed in LaCuO$_{2.64}$.

The frequency of the Cu1 line is constant a low $T$ but it acquires a marked temperature dependence (best described as linear) above 150~K (Fig.~\ref{cu1nqr}a). The date are consistent with an electronic crossover near 150~K but a continuous variation throughout the temperature range (such as an activated $T$ dependence shown in Fig.~\ref{cu1nqr}a with a characteristic energy $E\simeq$~620~K) cannot be fully excluded.

The spin-lattice relaxation rate $1/T_1$ of LaCuO$_{2.66}$ has similar values at 300~K as in LaCuO$_{2.64}$(Fig.~\ref{cu1nqr}). On cooling, however, the two compounds show a very different temperature dependence. In LaCuO$_{2.64}$, $1/T_1$ values remain high down to low $T$,~\cite{Walstedt94} consistent with a metallic ground state. In LaCuO$_{2.66}$, on the other hand, the particularly long $T_1$ of 14~s at $T=4.2$~K reveals that the Cu1 site is non-magnetic. This site should thus correspond to the non-magnetic Cu$^{3+}$ ions expected from the localization of holes at low temperature. In principle, Cu$^{2+}$ spins in dimer arrangements could also have a very small $1/T_1$ at low $T$ but they should have a different resonance frequency.

Above 150~K, the strong temperature dependence of $1/T_1$ can be fit to $1/T_1\propto \exp(-\Delta_{T1}/k_BT)$ with $\Delta_{T1}\simeq1216\pm14$~K (Fig.~\ref{cu1nqr}b,c). Within error bars, this gap value is equal to that inferred from the magnetic susceptibility, while being somewhat larger than that inferred from the resistivity (let us recall that the value of this latter might be affected by the grain boundaries). Since there is no doubt that the NMR signal comes from the bulk of the sample and is thus intrinsic, the comparable gap magnitudes tend to suggest that the activated dependence of the resistivity measured in this temperature range could indeed be intrinsic and not solely due to grain boundaries. Below 150~K, the deviation from the activated behavior indicates that the relaxation becomes dominated by residual relaxation processes (such as relaxation by paramagnetic impurities, quadrupole relaxation, etc.).

\subsection{$^{63,65}$Cu2a,b NQR}

In the 20-30~MHz range, two other Cu NQR sites (Cu2a and Cu2b) are found. Their intensity integrated over frequencies is weak, at most a few percents of the Cu1 intensity. At 4.2 K, the spin-lattice relaxation time $T_1$ of Cu2b is 24 ms, that is shorter than that of Cu1 by almost three orders of magnitude, indicating that, unlike Cu1, these sites carry a finite spin polarization. In agreement with the above analysis of the Curie-Weiss contribution to the spin susceptibility, these Cu2a and Cu2b sites should correspond to intrinsic paramagnetic spins 1/2 in the sample, rather than to diluted defects.

\subsection{$^{63,65}$Cu zero-field NMR signal}

At $T=1.6$~K and at frequencies around $f_0\simeq$121~MHz, a broad signal is attributed to Cu nuclei in the presence of an internal, static, magnetic field $H_{\rm int}=2\pi f_0/^{63}\gamma= 10.2$~T. Such a value of the resonance frequency is typical for AFM order in copper oxides, such as the high temperature superconductors. This signal is thus a zero-field NMR signal rather than a NQR signal since the dominant source of splitting of the nuclear levels is the Zeeman interaction rather than the quadrupolar one.
The presence of frozen moments on some of the Cu sites of the sample agrees with a muon spin rotation ($\mu$SR) study performed on the same sample,\cite{Mendels04} which indicates that AFM order is present below 40-50~K in $\sim$20\% of the sample volume. Confirmation of this number was not possible from the present nuclear resonance data. This signal was found to disappear very rapidly on warming (it was much weaker at 4.2~K) most probably because of dramatic shortening of the transverse relaxation time $T_2$. The NQR signal from these sites remains to be observed in the paramagnetic phase.

\subsection{ESR}
\begin{figure}
\centerline{\includegraphics[width=3.4in]{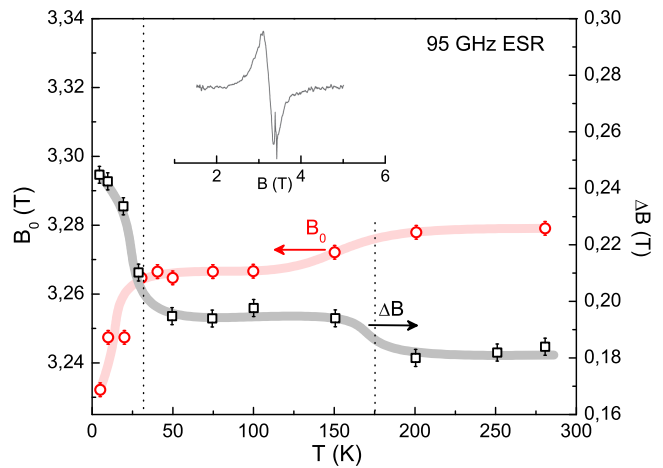}}
\caption{(Color online) Field $B_0$ and width $\Delta B$ of the ESR resonance line (shown in the inset for $T=5$~K) at a frequency of 95~GHz. The width is defined as the peak-to-peak distance of the signal shown in the inset. Vertical dotted lines show temperature crossovers.}
\label{ESR}
\end{figure}

The ESR line at 95~GHz is a single Lorentzian, indicating that any anisotropy of the $g$ factor is not resolved at this frequency. Its intensity at low temperature (not shown) follows a Curie-Weiss law suggesting that this signal stems from the same spins as probed in the bulk susceptibility. As Fig.~\ref{ESR} shows, the temperature dependence of both the field and the width of the resonance shows a first crossover near 150-180~K followed by a sharper change near 30-40 K (for each of these two temperature ranges, the lower $T$ value is that indicated by the resonance field data, while the larger one is given by the width data). The higher temperature crossover should be related to the change observed in NQR frequency in the same temperature range. The additional shift/broadening of the resonance field below 30-40~K is clearly due to the magnetic freezing found to onset around 40-50~K in $\mu$SR.\cite{Mendels04} These preliminary ESR results demonstrate that the ESR signal is from intrinsic paramagnetic spins in LaCuO$_{2.66}$.

\begin{figure}[t!]
\centerline{\includegraphics[width=3.3in]{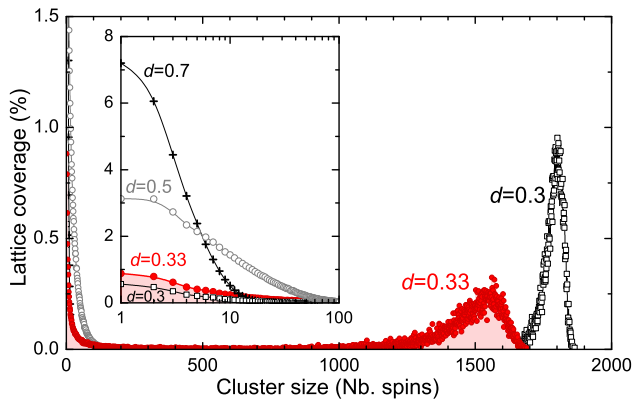}}  \caption{(Color online) Coverage of the lattice (= Number of clusters of a given size times the number of sites in the cluster divided by the total number of lattice sites), as a function of cluster size ({\it i.e.} number of sites in a magnetic cluster), for different values of the dilution level $d$. The inset shows an expanded view at low values of cluster sizes. The data are from simulations of a planar lattice of 2700 sites. Since no interplanar exchange coupling is considered, the lattice consists of a single plane.} 
\label{histo}
\end{figure}

\section{Numerical simulations}

\subsection{Cluster sizes}

The experimental evidence for localization of the doped holes at low temperature signifies that LaCuO$_{2.66}$ should be described as a 1/3-diluted kagome lattice. We thus performed a number of numerical simulations in order to provide a concrete picture of the system at such a high dilution level $d=0.33$, which turns out to almost correspond to the site percolation threshold $p_c$=0.6527 of the kagome lattice~\cite{Henley} ({\it i.e.} $d_c=1-p_c=0.3473$). The distribution of magnetic cluster size~(Fig.~\ref{histo}) was obtained by averaging over a large number of randomly depleted lattices and it was checked to converge rapidly when increasing the lattice size. At large dilution levels, only small clusters (typically less than 10 sites) are present. For intermediate dilution $d\simeq0.5$, there is a homogeneous distribution of sizes up to typically 100 sites (see inset of Fig.~9).

As the percolation threshold $p_c=1-d_c$ is crossed, the intermediate-size clusters appear to coalesce into large clusters and the distribution becomes essentially bimodal with a minority of sites in small clusters and a large majority of sites embedded in extended clusters, some of which are percolating. An example of a diluted kagome lattice of $N=1200$ sites at $d=0.33$ is shown in Fig.~\ref{percol}. At this concentration and for $N=2700$, clusters of less than 100 sites cover about 12~\% of the lattice while clusters of more than 1000 sites cover 55~\%.

\subsection{Cluster morphologies}

\begin{figure*}
\centerline{\includegraphics[width=5.3in]{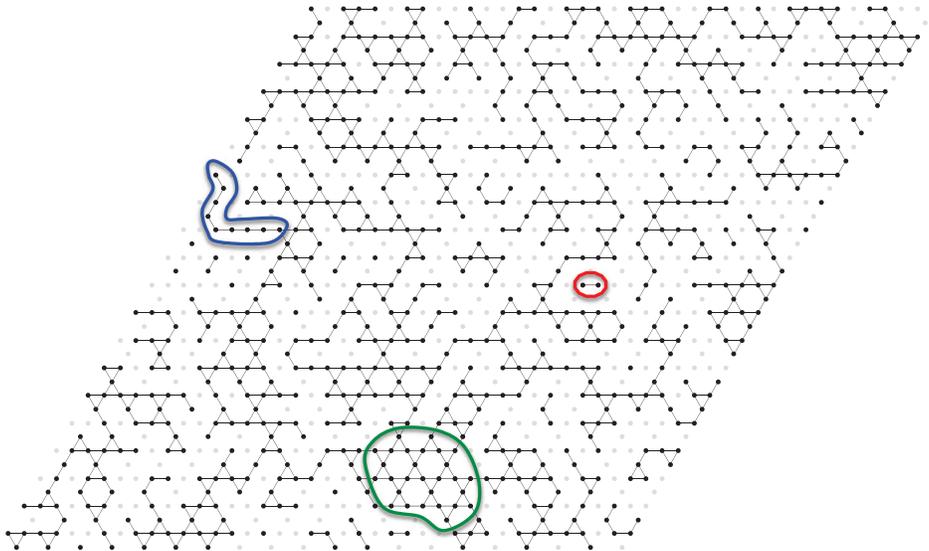}}
\vspace{-1cm}
\caption{(Color online) Example of cluster distribution from a 33\% random site dilution on the kagome lattice with $20\times20\times3=1200$ planar sites with three sites per unit cell. Examples of a small cluster (red line), and of a non-frustrated environment (blue line) and of a frustrated environment (green line) in the same big cluster are shown. Not fully apparent in this figure is the fact that clusters of less than 100 sites cover approximately 12~\% of the lattice while clusters of more than 1000 sites cover $\sim$55~\%.} 
\label{percol}
\end{figure*}

Fig.~\ref{percol} shows that, to first approximation, two very different types of connectivity can be singled out in the magnetic clusters. On the one hand, there are spins for which the local environment remains identical to the pristine kagome lattice. On the other hand, there are spins in a weakly frustrated environment, including chain-like fragments. We hypothesize here
that the weakly frustrated spins will have a tendency towards AFM-like ordering whereas the more frustrated sites will still form spin liquid-like arrangements (which may be of various kinds: valence bond crystals, dimer liquid, valence bond glass, etc.) and thus remain mostly non-magnetic with a susceptibility relatively close to that of the pure kagome lattice.

Furthermore, a similar description has already been proposed to explain the existence of two different sites in the M\"{o}ssbauer spectra of LuFeMgO$_4$, a triangular lattice of Fe$^{3+}$ spins with a site dilution level of 50 \% coinciding with the percolation threshold of this lattice.\cite{Ikeda} The magnetic freezing of this compound (around 33~K, that is well below the Curie-Weiss temperature of -500~K) has thus been ascribed to the freezing of the weakly frustrated spins. Although the ground state of the triangular lattice (long-range ordered with spins oriented at 120$^{\circ}$ from each other) is different from that of the kagome lattice (non-magnetic with short-ranged spin correlations), there are obvious analogies with the situation in LaCuO$_{2.66}$.

Of course, this description is a caricature of the actual situation since there exist many different morphologies with all possible intermediate levels of frustration in between these two limits. Nonetheless, this simple description helps to understand qualitatively the inhomogeneous magnetism in LaCuO$_{2.66}$ and a more sophisticated model is beyond the scope of this paper. Note also, as the next section will emphasize, that the local environment and the symmetry of sites belonging to large clusters is irrelevant from the point of view of the susceptibility.

\subsection{Calculation of the magnetic susceptibility}

\begin{figure}[t!]
\centerline{\includegraphics[width=3.3in]{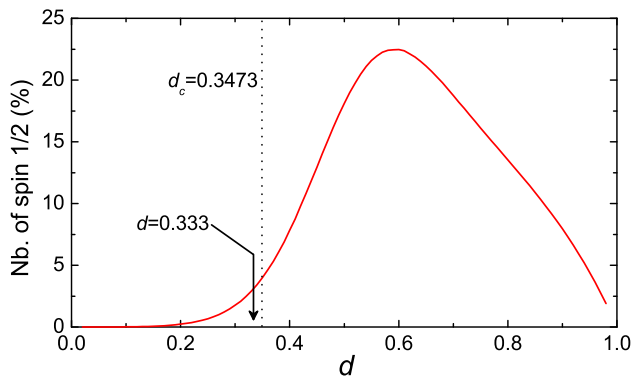}}
\caption{(Color online) Calculated percentage of Cu sites contributing one spin 1/2 to the Curie-Weiss susceptibility, as a function of the dilution level per Cu site $d$. The vertical line at $d_c=0.3473$ indicates the site percolation threshold $p_c$=0.6527 of the kagome lattice ($d_c=1-p_c$). See text for details.}
\label{calcchi}
\end{figure}

To first approximation, the Curie-Weiss susceptibility arises from isolated $S=1/2$ sites and from orphan spins~\cite{Schiffer,Moessner} within clusters with an odd number of sites. In fact, it can be shown that the largest clusters ({\it i.e.} those with the most pronounced kagome character) have a negligible contribution to the susceptibility in the temperature range of interest here as their Curie-like tail becomes non-negligible only at very low $T$. A practical cutoff value of 13 was chosen for the size of the clusters considered in the calculation of the susceptibility. Each of these clusters of $N$ sites ($N\leq13$) was approximated to be a chain of $N$ sites, whose magnetic linear susceptibility was then calculated from exact diagonalization. The linear topology for the small clusters is obviously a crude approximation. In particular, there is a non linear inverse susceptibility yielded by the progressive construction of collective spin 1/2 resulting from entanglement in small clusters with triangular geometry.\cite{Isoda,Robert,Azuma} However at low T, the entangled spins 1/2 should yield a Curie-Weiss behavior. Finally, an average susceptibility was calculated by considering the distribution of small clusters in a set of randomly depleted kagome lattices, for each dilution level. The susceptibility computed with these assumptions was fitted with a Curie-Weiss model. The resulting percentage of spins 1/2 per total number of Cu is shown in Fig.~\ref{calcchi} as a function of $d$. For 1/3 dilution, the value of $\approx$3~\% matches well the experimental result of 5~\% (which also includes the contribution from extrinsic defects) (Section IV.B).

\section{Summary of main results}

\subsection{From experiments}

As far as magnetic properties are concerned, our experimental results can be summarized in terms of coexistence of different Cu sites at low temperature in LaCuO$_{2.66}$:

(1) Cu$^{2+}$ sites experiencing magnetic order. We conjectured that these sites, which are observed in nuclear resonance and $\mu$SR experiments,\cite{Mendels04} belong to extended parts of the clusters, such as chain fragments, whose geometry effectively releases the frustration. 20~\% of the implanted muons experience a static field from these moments, suggesting that the magnetic sites represent a fraction of the sites which is of the order of 20~\%.\cite{Mendels04}

(2) Isolated paramagnetic Cu$^{2+}$ spins. These sites can account for the Curie-like tail of the susceptibility, possibly for the Cu2a and Cu2b NQR sites, and for the Schottky anomaly in the specific heat. We have argued that they originate from small clusters, although some of them may also be of extrinsic origin.

(3) Non-magnetic Cu$^{3+}$. All, or part, of the Cu1 NQR line must come from these Cu$^{3+}$ sites. Given the resistivity and specific heat evidence of a fully insulating ground state, they should represent 33~\% of the total number of sites.

It should be noted that it was not possible to single out the spin-singlet Cu$^{2+}$ sites which are expected to be present in large clusters where the local environment remains identical to the pristine kagome lattice as well as in small clusters with an even number of spins (such as shown in Fig.~\ref{percol}).

The insulating ground state of LaCuO$_{2.66}$ below $\sim$150~K is in agreement with all measurements and especially with the high values of the resistivity and the absence of a linear-in-$T$ contribution to the specific heat. Above $180\pm20$~K, however, the resistivity, the bulk magnetization and the spin-lattice relaxation of the Cu1 site, all show an activated behavior. This is reminiscent of semiconductors for which the thermal activation arises from the thermal excitations of carriers in the valence band. Around 180~K, there is also a crossover in the position and the width of the absorbtion line in electron spin resonance.

\subsection{From numerical simulations}

We have seen that, for a dilution level as large as 33~\%, the identity of the original kagome lattice is seriously degraded and, despite remaining patches of local kagome character, the system becomes mostly an inhomogeneous pattern of magnetic clusters. Therefore, the peculiar properties found for weak dilution in $S=1/2$ kagome systems (each impurity binds strong spin singlets in its vicinity~\cite{impurities}) are unlikely to be relevant for LaCuO$_{2.66}$. Instead, the properties of each spin next to a Cu$^{3+}$ will primarily be determined by the morphology of the part of the cluster it belongs to. This does not necessarily mean that all memory of the kagome physics is lost. For example, in the $S=3/2$ kagome-bilayer compound SCGO, remarkable properties associated with the spin liquid state (such as the original magnetic excitations responsible for a $T^2$ contribution to the specific heat or the peculiar dynamics probed by muon spectroscopy~\cite{Ramirez92,Keren}) are still observed at a dilution level as high as 60 \%, {\it i.e.} above the percolation threshold $p_c$=0.5.\cite{Henley}

\section{Discussion}

\subsection{Origin of the partial magnetic freezing}

First, it must be realized that the partial magnetic freezing does not arise from parasitic phases or from strong inhomogeneity of the sample: the former is ruled out by (X-ray and neutron) diffraction data~\cite{Garlea03} and the latter is inconsistent with the observation of well-defined NQR lines and of a well-defined structure (recall that oxygen dopants are not randomly located, they are well-ordered in Cu planes where they build the kagome lattice itself). Also, the well-defined oscillation seen in $\mu$SR for both LaCuO$_{2.66}$ and YCuO$_{2.66}$ (with an even larger magnetic fraction for the latter) speak in favor of an intrinsic magnetism from macroscopic-size clusters.\cite{Mendels04} Thus, the frozen magnetic fraction is an intrinsic property of LaCuO$_{2.66}$.

Even if the two-dimensional, frustrated and quantum nature of the kagome lattice of spins 1/2 is unfavorable to long range ordering~\cite{Reviews}, the occurrence of magnetic freezing in LaCuO$_{2.66}$ is not really a big surprise: inhomogeneous magnetism with partial fractions of frozen spins is already found in relatively pure compounds approaching the perfect kagome geometry.\cite{volbor,Quilliam11} Part of the explanation probably lies in the presence of the Dzyaloshinskii-Moriya interaction, which is known to drive the kagome antiferromagnet towards magnetic ordering,\cite{Elhajal} even for the quantum spins $S=1/2$.\cite{Cepas}

Dzyaloshinskii-Moriya interaction is also expected for the non centrosymmetric Cu-O-Cu bonds in LaCuO$_{2.66}$ but since its magnitude is not known, it is difficult to assess its role in the spin freezing here. In any event, we expect that the destructuralization of the kagome lattice at 1/3-dilution should still play a dominant role. The hypothesis that magnetic order is nucleated at those sites for which the local environment is mostly non-frustrated (chain-like for instance) readily justifies the partial character of the frozen phase. This scenario of locally released frustration also provides a natural boost of the onset temperature of magnetic freezing which is needed to explain values as high as 40-50~K in LaCuO$_{2.66}$, that is equivalent to at least $J$/10 (see discussion on $J$ in section~III.B).

In any case, it is likely that magnetic order in these frustrated compounds arises from the conjunction of different factors and not from a single one. For example, the probable inter-triangle exchange interactions in YCuO$_{2.66}$ represent a deviation from the simple nearest-neighbor kagome model which lowers the degree of frustration and thus favors spin ordering. This presumably explains why the magnetic volume fraction in YCuO$_{2.66}$ is as high as 80~\%,\cite{Mendels04} much higher than 20~\% in LaCuO$_{2.66}$, for the same dilution level. However, orbital-ordering and/or Jahn-Teller effects might have to be considered as well.

\subsection{Localization of doped holes}

The properties of LaCuO$_{2.66}$ appear different from those of LaCuO$_{2.64}$. LaCuO$_{2.66}$ seems to have an insulating ground state with magnetic order on part of the sites, while LaCuO$_{2.64}$ apparently has features (such as a $T$-linear term in $C_p$ and a single NQR site with short relaxation time $T_1$) which suggest a metallic ground state without any magnetic order.~\cite{Cava93,Ramirez94,Walstedt94} Because the perfect kagome lattice is realized only for $\delta=0.66$, the small difference in the nominal planar-O content of the two compounds implies a difference in their electronic structure which can be much more effective than the difference in hole-doping in producing different ground states. Furthermore, given the lack of a well-defined planar structure for $\delta=0.64$, there is a much larger uncertainty on the oxygen content for this compound. Therefore, the two compounds may be significantly more different than what their similar formulas suggest. Nevertheless, the striking difference in physical behaviors is unlikely to be simply explained in terms of structural differences. Indeed, the larger disorder in LaCuO$_{2.64}$ should make the sample more insulating, while it is observed to be more metallic than LaCuO$_{2.66}$.

A possible solution to this paradox is that the commensurate value of the oxygen concentration $\delta=2/3$ locks the otherwise metallic system into an insulator. A scenario of charge ordering at $\delta=2/3$ is appealing for rationalizing a commensurability lock-in and for partially relieving frustration.\cite{Dolores96} It might also be compatible with the crossover detected near 150-180~K in the NQR and ESR measurements (Fig.\ref{cu1nqr}a). However, we only mention this scenario here as a possibility: in the absence of any direct evidence, charge order remains a speculation here and crossovers could as well be produced by hole localization at random locations.

\section{Conclusion}

We have found that LaCuO$_{2.66}$ has an insulating ground state which makes it a realization of the 1/3 depleted kagome lattice. We have found inhomogenous magnetism which numerical simulations suggest to arise from the heterogeneous distribution of cluster sizes and morphologies near the site-percolation threshold. Of course, a lot remains to be done to understand these complex delafossite compounds. Although apparently not displaying the metallic properties of sodium cobaltates Na$_x$CoO$_2$ (a doped triangular lattice), this class of copper oxides appears as an uncharted playground for studying the doping of a frustrated lattice and would be worth deeper theoretical and experimental investigations.

\section{Acknowledgements}

We are grateful to F. Bert, C. Berthier, C. Lacroix, H. Mayaffre, P. Mendels and D. Nunez-Regueiro for enlightening discussions and to Y. Berthier for assistance in the NQR experiments. We also thank T. Grenet, J. Delahaye and F. Gay for help and advice concerning the resistivity measurements. VOG acknowledges the support by the Scientific User Facilities Division, Office of Basic Energy Sciences, U.S. Department of Energy.

\end{document}